\documentstyle[aps,12pt]{revtex}
\newcommand{\bb}{\begin{equation}}
\newcommand{\ee}{\end{equation}}
\newcommand{\ba}{\begin{array}}
\newcommand{\ea}{\end{array}}

\begin {document}
\baselineskip 15pt 

\title{Pair creation of neutral particles in a vacuum by external
        electromagnetic fields in 2+1 dimensions
        \thanks{published in J. Phys. G {\bf 25} (1999) 1793-1795.
        \copyright IOP publishing Ltd. 1999}}
\author{Qiong-gui Lin\thanks{E-mail addresses: qg\_lin@163.net,
stdp@zsu.edu.cn}}
\address{China Center of Advanced Science and Technology (World
	Laboratory),\\
        P.O.Box 8730, Beijing 100080, People's Republic of China\\
        and\\
        Department of Physics, Zhongshan University, Guangzhou
        510275,\\
        People's  Republic of China \thanks{Mailing address}}

\maketitle
\vfill

\begin{abstract}
{\normalsize
Neutral fermions of spin $\frac 12$ with magnetic moment can interact
with electromagnetic fields through nonminimal coupling.
In 2+1 dimensions the electromagnetic field strength plays the same
role to the magnetic moment as the vector potential to the
electric charge. This duality enables one to obtain physical results
for neutral particles from known ones for charged particles.
We give the probability of neutral particle-antiparticle pair creation
in the vacuum by non-uniform electromagnetic
fields produced by constant uniform charge and current densities.}
\end{abstract}
\vfill
\leftline {PACS number(s): 03.70.+k, 11.10.Kk, 11.15.Tk}
\newpage

Pair creation of charged particles in the vacuum by an external
electric field was first studied by Schwinger several
decades ago [1]. Related problems have been discussed by
many authors, for example, in Refs. [2-6].
Pure magnetic fields do not lead to pair creation. However, inclusion
of a magnetic field changes the result for a pure electric field.
This has been studied by several authors [7-11]. The probability of
pair creation in the vacuum was also calculated in lower dimensions
[11] and in more general dimensions [8]. Similar 
subjects are now 
widely discussed in string and black hole theory.

In this paper we extend the subject to neutral particles of spin
$\frac 12$. Neutral particles with magnetic moment can interact
with electromagnetic fields through nonminimal coupling. The
Aharonov-Casher effect [12] is a well known indication for this
interaction, and has been observed in experiment [13]. So we expect
that neutral particle-antiparticle pairs may be created in the
vacuum by external electromagnetic fields.
In 2+1 dimensions this can be verified by explicit calculations.
For certain electromagnetic fields
the probability of pair creation can be calculated exactly.
Indeed, the result can be obtained from that for charged particles
by using the duality between charged and neutral particles,
respectively in external vector potentials and
electromagnetic fields.
Unfortunately, the problem is
rather complicated in 3+1 dimensions, and useful results are still
not available.

Consider a neutral fermion of spin $\frac 12$ with mass $m$ and
magnetic moment $\mu_m$, moving in an external electromagnetic field.
The equation of motion for the fermion field $\psi$ is [14]
\bb
(i\gamma\cdot\partial-\mu_m\sigma\cdot F/2-m)\psi=0,
\ee            
where $\gamma\cdot\partial=\gamma^\mu\partial_\mu$, $\sigma\cdot F
=\sigma^{\mu\nu}F_{\mu\nu}$, $F_{\mu\nu}$ is the field strength of the
external electromagnetic field, and
\bb
\sigma^{\mu\nu}={i\over 2}[\gamma^\mu, \gamma^\nu].
\ee               
In the following we work in 2+1 dimensions. The Dirac matrices
$\gamma^\mu$ satisfy the relation
\bb
\gamma^\mu\gamma^\nu=g^{\mu\nu}\mp i\epsilon^{\mu\nu\lambda}
\gamma_\lambda,
\ee     
where $g^{\mu\nu}={\rm diag}(1, -1, -1)$, $\epsilon^{\mu\nu\lambda}$
is totally antisymmetric in its indices and $\epsilon^{012}=1$.
The minus sign in (3) corresponds to representations equivalent to
the one
$$
\gamma^0=\sigma^3,\quad \gamma^1=i\sigma^1,\quad \gamma^2=i\sigma^2,
\eqno(4a)
$$
and the plus sign corresponds to representations equivalent to
the one
$$
\gamma^0=-\sigma^3,\quad \gamma^1=i\sigma^1,\quad \gamma^2=i\sigma^2,
\eqno(4b)
$$
which is inequivalent to (4a). The difference between the two cases
would not affect physical results. So we will work with the first
case [with the minus sign in (3)]. The relation (3) is of crucial
importance for the following discussion. Note that in other space-time
dimensions there is no similar relation. Using (3) we have
$\sigma^{\mu\nu}=\epsilon^{\mu\nu\lambda}\gamma_\lambda$. We define
\addtocounter{equation}{1}
\bb
f^\mu=\frac 12 \epsilon^{\mu\alpha\beta}F_{\alpha\beta},
\ee       
and have
\bb
\sigma\cdot F/2=\gamma\cdot f.
\ee             
Then (1) becomes
\bb
[\gamma\cdot(i\partial-\mu_m f)-m]\psi=0.
\ee            
Compared with the equation of a charged fermion with spin $\frac 12$
and charge $q$ moving in an external electromagnetic vector potential
$A_q^\mu$
\bb
[\gamma\cdot(i\partial-qA_q)-m_q]\psi_q=0
\ee            
where a subscript $q$ is used to indicate the charged particle, one
easily realize that $f^\mu$ plays the same role to the neutral
particle as $A_q^\mu$ to the charged one. This duality has been
noticed in the study of the Aharonov-Casher effect [12,13].
Consequently, constant uniform fields $F_{\mu\nu}$ have no physical
effect on neutral particles. To create particle-antiparticle pairs
in the vacuum, non-uniform fields are necessary. Because of the above
duality, we call $f^\mu$ the dual vecter potential, and
$f_{\mu\nu}=\partial_\mu f_\nu-\partial_\nu f_\mu$ the dual field
strength. The dual magnetic field is
$$
b=-f_{12}=\partial_\lambda F^{\lambda 0}=\nabla\cdot{\bf E},
\eqno(9a)$$
and the dual electric field is
$$
e_x=f_{01}=-\partial_\lambda F^{\lambda 2}=\partial_t E_y+
\partial_x B,
\eqno(9b)$$
$$
e_y=f_{02}=\partial_\lambda F^{\lambda 1}=-\partial_t E_x+
\partial_y B.
\eqno(9c)$$
Using the Maxwell equation $\partial_\lambda F^{\lambda\mu}=J^\mu$
where $J^\mu$ is the external source producing the field
$F_{\mu\nu}$, we have
$$
b=J^0\equiv\rho, \eqno(10a)$$
$$
e_x=-J^2\equiv -J_y, \quad e_y=J^1\equiv J_x.
\eqno(10b)$$
This means that the current $(-J_y, J_x)$ plays the same role to the
neutral particles as an electric field to the charged ones,
and the charge density $\rho$ plays the same role as a magnetic field.
We can now use the conclusions and results of [11] without further
calculations. Consider an external electric charge density $\rho$
and an electric current density ${\bf J}$, both being
constant and uniform. The field strength $F_{\mu\nu}$ produced by
them is not uniform.
If $|{\bf J}|<|\rho|$ there is no pair creation. When
$|{\bf J}|>|\rho|$, we have the probability, per unit time and per
unit area, of neutral particle-antiparticle pair creation in the
vacuum
\addtocounter{equation}{2}
\bb
w={(|\mu_m|{\cal J})^{3\over2}\over 4\pi^2}\sum_{n=1}^\infty
{1\over n^{3\over2}}
\exp\left(-{n\pi m^2\over |\mu_m|{\cal J}}\right),
\ee                   
where
\bb
{\cal J}=\sqrt{{\bf J}^2-\rho^2}.
\ee             
$w$ is independent of the space-time position as both $\rho$ and
${\bf J}$ are constant and uniform. Also note that the result is
Lorentz invariant.

In conclusion, neutral particle-antiparticle pairs can be created
in the vacuum by external electromagnetic fields in 2+1 dimensions,
if these particles have nonvanishing magnetic moment. For
electromagnetic fields produced by constant uniform sources, the
probability of pair creation in the vacuum can be obtained from
similar results for charged particles. The crucial point is the
duality between magnetic moment and electric charge. It is expected
that similar conclusions can be achieved in 3+1 dimensions.
Unfortunately,
in 3+1 dimensions there is no relation similar to (3), and the
duality used here does not exist. Consequently, the calculations
in 3+1 dimensions involve considerable mathematical difficulties.
We cannot work out an explicit nontrivial result even for a field
configuration similar to that in 2+1 dimensions, i.e., a field
configuration where $B_x=B_y=E_z=0$ and everything is independent of
$z$. At present, we can only calculate the simplest case for a
constant uniform electric field, but the result turns out to be
trivial. For a constant uniform magnetic field, the calculations can
also be carried out
analytically, but the final result involves divergent integrals which
we still do not know how to regularize.
It seems that some other techniques
should be developed to deal  with the subject in 3+1 dimensions.

\vskip 1cm

This work was supported by the
National Natural Science Foundation of China.

\end{document}